\documentclass[12pt,draftcls,onecolumn]{IEEEtran}

\usepackage{bm}
\usepackage{amssymb,amsmath}
\usepackage[ruled,vlined]{algorithm2e}
\usepackage{float,graphicx}
\usepackage[acronym,shortcuts]{glossaries}
\usepackage{comment}
\usepackage{color}
\usepackage{soul}
\usepackage[export]{adjustbox}

\newacronym{AF}{AF}{amplify and forward}
\newacronym{AN}{AN}{atomic norm}
\newacronym[plural=AoAs,firstplural=angles of arrival (AoAs)]{AoA}{AoA}{angle of arrival}
\newacronym{AoD}{AoD}{angle of departure}
\newacronym{AWGN}{AWGN}{additive white Gaussian noise}
\newacronym{BS}{BS}{base station}
\newacronym{CDF}{CDF}{cumulative distribution function}
\newacronym{CF}{CF}{compute and forward}
\newacronym{CSI}{CSI}{channel-state information}
\newacronym{D2D}{D2D}{device-to-device}
\newacronym{DF}{DF}{decode and forward}
\newacronym{FD}{FD}{full-duplex}
\newacronym{GAE}{GAE}{gridless angle estimation}
\newacronym{GEC}{GEC}{gridless estimation and clustering}
\newacronym{HD}{HD}{half-duplex}
\newacronym{ID}{ID}{ideal}
\newacronym{iid}{i.i.d.}{independent and identically distributed}
\newacronym{LS}{LS}{least square}
\newacronym{OFDMA}{OFDMA}{orthogonal frequency division multiple access}
\newacronym{mmWave}{mmWave}{millimeter-wave} 
\newacronym{MAC}{MAC}{multiple access channel}
\newacronym{MIMO}{MIMO}{multiple-input multiple-output}
\newacronym{MIMOME}{MIMOME}{MIMO multiple-Eve}
\newacronym{MSE}{MSE}{mean square error}
\newacronym{MMSE}{MMSE}{minimum \ac{MSE}}
\newacronym{MT}{MT}{mobile terminal}
\newacronym{MUSIC}{MUSIC}{multiple signal classification}
\newacronym{NOMA}{NOMA}{non-orthogonal multiple access}
\newacronym{nLOS}{nLOS}{no-line-of-sight}
\newacronym{PBCE}{PBCE}{projection-based correlation estimation}
\newacronym{PDF}{PDF}{probability density function}
\newacronym{PLS}{PLS}{physical layer security}
\newacronym{RB}{RB}{random binning}
\newacronym{ROT}{ROT}{raise-over-thermal}
\newacronym{SCF}{SCF}{scaled compute-and-forward}
\newacronym{SGE}{SGE}{subspace gridless estimation}
\newacronym{SIC}{SIC}{successive interference cancellation}
\newacronym{SOP}{SOP}{secrecy outage probability}
\newacronym{SVD}{SVD}{singular value decomposition}
\newacronym{SNR}{SNR}{signal to noise ratio}
\newacronym{TDD}{TDD}{time-division duplexing}
\newacronym{UE}{UE}{user equipment}

\usepackage[normalem]{ulem}
\usepackage[user]{zref}
\newcommand{\sot}[1]{}

\newcounter{revc}
\makeatletter \zref@newprop{revcontent}{} \zref@addprop{main}{revcontent}
\zref@newprop{revsec}{} \zref@addprop{main}{revsec}
\zref@newprop{revpage}{} \zref@addprop{main}{revpage}

\newcommand{\revi}[2]{
\zref@setcurrent{revsec}{\thesection}%
\zref@setcurrent{revpage}{\thepage}%
\zref@setcurrent{revcontent}{#2}%
\refstepcounter{revc}%
\label{#1}%
\zlabel{#1}%
\textcolor{blue}{#2}%
}

\makeatletter
\newcommand\revitext[2]{
  \immediate\write\@auxout{\unexpanded{\global\long\@namedef{mytext@#1}{#2}}}%
  \textcolor{black}{#2}%
}

\newcommand\revptext[1]{%
  \textcolor{black}{\ifcsname mytext@#1\endcsname
    \@nameuse{mytext@#1}} %
  \else
    ``??''
  \fi
}

\makeatother

\begin{document}
\sloppy

\title{Estimation of Interference Correlation \\ in mmWave Cellular Systems}

\author{Stefano Tomasin$^1$, Raphael Hasler$^1$, Antonia M. Tulino$^2$, and Matilde Sánchez-Fernández$^3$ \\
\small $^1$: Universit\`a degli Studi di Padova, Italy \\
$^2$: Universit\`a degli Studi di Napoli Federico II, Naples, Italy\\
$^3$: Universidad Carlos III de Madrid, Spain. \\
Corresponding author: Stefano Tomasin, stefano.tomasin@unipd.it
}

\maketitle

\begin{abstract}

We consider a cellular network, where the uplink transmissions to a \ac{BS} are interferenced by other devices, a condition that may occur, e.g., in cell-free networks or when using \ac{NOMA} techniques. Assuming that the \ac{BS} treats this interference as additional noise, we  focus on the problem of estimating the interference correlation matrix from received signal samples. We consider a \ac{BS} equipped with multiple antennas and operating in the \ac{mmWave} bands and  propose techniques exploiting the fact that channels comprise only a few reflections at these frequencies. This yields a specific structure of the interference correlation matrix that can be decomposed into three matrices, two rectangular depending on the \ac{AoA} of the interference and the third square with smaller dimensions. We resort to gridless approaches to estimate the \acp{AoA} and then project the \acl{LS} estimate of the interference correlation matrix into a subspace with a smaller dimension, thus reducing the estimation error. Moreover, we derive two simplified estimators, still based on the gridless angle estimation that turns out to be convenient when estimating the interference over a larger number of samples. 
\end{abstract}


\glsresetall 

\section{Introduction}

The evolution of cellular networks is pushing towards \ac{NOMA} schemes, wherein the wireless resource is not exclusively assigned to a user (as in orthogonal multiple access) but is shared among several devices that interfere. Such an approach may be applied either within each cell (in what is more specifically \ac{NOMA}, \cite{9502643}) or among different cells, e.g., in dynamic \ac{TDD} networks \cite{9139384} and is particularly attractive in systems with partial or distributed coordination, such as in \ac{D2D} communications \cite{9632534} or cell-free networks \cite{9650567}. \revitext{intasnoise1}{In all these cases, the interference among devices becomes challenging and is addressed by several techniques that can be classified according to the underlying interference model: when interference is seen as an additional data signal with a suitable structure (e.g., with known modulation and coding parameters) \ac{SIC} or multiuser decoding solutions are applied; when interference is seen as additional noise, it is handled by suitable signal processing at the transmitter or the receiver, e.g., by beamforming and combining signals over multiple antennas in \ac{MIMO} systems. The first approach, which is preferable under strong interference coming from a few users, requires the estimation of the interference channels, and yields intensive computations. The latter approach is preferable when several low-power interferers are present, as they cannot be demodulated without errors; this approach  requires the estimation of the statistics of the interference and entails fewer computations.} \revitext{rev32}{However, interference can always be treated as additional noise, even under strong interference, thus our approach has a very wide application. This is particularly true when either strong interferers cannot be decoded, e.g., because  the receiver either does not know the used modulation or coding or does not have enough computational capabilities. In all these cases, the estimation of the interference correlation matrix turns out to be very useful.}

We focus on the latter scenario, where interference is treated as additional noise, and consider the problem of estimating its correlation matrix when the receiver is equipped with multiple antennas. The estimation of the interference statistics is not only useful for the design of beamformers but also to improve the channel estimation \cite{6415397}, estimate the phase noise and frequency offset  \cite{1677918}, and perform link adaptation \cite{doi:https://doi.org/10.1002/9780470978504.ch10}. \revitext{covmatrix}{Note however that we aim at estimating the correlation matrix of the interference averaging over the signal transmitted by the interferers, but assuming their channel as time-invariant: this makes our problem significantly different from the  covariance matrix estimation of the channel, where the average is taken with respect to the channel fading (see for example \cite{8338075}).}

Several works in the literature have addressed the estimation of the noise variance, under the assumption of independent identically distributed noise samples on each receive antenna. Among other works, we mention \cite{8082536}, where the noise variance estimation is considered in the context of massive \ac{MIMO} assuming uncorrelated noise at the antennas and under practical impairments such as pilot contamination. Although a different noise power is assumed at each antenna, noise samples at different antennas are still considered uncorrelated. In this paper, we instead focus on the case wherein interference is correlated among the various antennas. \revitext{rev11}{In this respect, in the literature we can find exponential models for the correlation of the noise among antennas, see \cite{6965901} and references therein. These models neither exemplify any particular propagation scenario nor are suited to capture the underlying composite interference signals. In fact, they are typically proposed as the simplest analytical model that can be easily adapted to empirically derived covariances \cite{exponential_model}. Nevertheless, exponential correlation modeling has some drawbacks: on one hand, the fitting may not be straightforward given that they are not linear functions, on the other, they may miss relevant components, e.g., at high frequencies.}
In \cite{4133041, 4803746} the training sequence for channel estimation in \ac{MIMO} systems with colored noise is designed, but no specific solution is proposed for the estimation of the noise correlation matrix. In \cite{8553218} a generic covariance matrix of the interference is considered, and the problem of its joint estimation with the channel is addressed and solved by a low-rank approximation of the channel. However, no particular structure of the interference correlation matrix is assumed. The estimation of colored interference is considered in \cite{9139384} and a \ac{LS} technique is proposed. The noise correlation matrix estimation is included in the estimation of the channel correlation in \cite{dietrich2006estimation}, and a modified version of the \ac{LS} estimator is derived, taking into account the channel statistics. 

\revitext{rev12}{However, to our knowledge, there are no specific solutions for interference correlation matrix estimation when operating at \ac{mmWave} or Terahertz (THz) bands. Indeed, at those frequencies,  the channels exhibit only a few propagation paths, due to the strong attenuation incurred by radio signals. This feature has been extensively exploited to obtain accurate channel estimators
\cite{9165822}.  Still, it remains not used for the estimation of the interference correlation matrix in systems operating at \ac{mmWave} or THz bands.} \revitext{rev13}{Note that, although the presence of few paths could make the beamforming to the interfering users more accurate, the strong attenuations combined to more flexible architectures (including \ac{D2D}, cell-free, and dynamic-TDD communications) may make the detection and cancellation of {\em all} interfering users more problematic. Thus, also with \acp{mmWave} the estimation of the interference correlation matrix is still relevant.}
 
In this paper, we propose three techniques to estimate  the interference correlation matrix in \ac{mmWave} or THz systems with different complexity and performance. All solutions exploit the structure of the interference channels, and the objective is to obtain more accurate estimations using fewer samples than those needed by the \ac{LS} approach. In particular, we consider a receiver equipped with a linear array of antennas and  first observe that the interference correlation matrix can be decomposed into the product of three matrices, where the two external depend only on the \acp{AoA} of the interference signals, while the internal one represents the correlation   among the interference signals. The internal matrix has  a much lower rank than the interference correlation matrix since it only provides the correlation of signals with the (few) different \acp{AoA}. Then, we propose a procedure by which we estimate a) the \acp{AoA}, b)  the internal correlation matrix, and c)  the interference matrix. Lastly, we reconstruct the interference correlation matrix.

In particular, for the \ac{AoA} estimation, we adopt a \ac{GAE} approach  \cite{9380368}, where the estimation is described as a gridless identification of several multi-dimensional frequency vectors, each corresponding to a different \ac{AoA}. Although this estimator shows an improvement over the \ac{LS} estimator already when using a few samples for each antenna, it becomes infeasible when applied on several samples. Therefore, we introduce two alternative approaches, still based on \ac{GAE}: the \ac{SGE} and the \ac{GEC}. \ac{SGE} first estimates the subspace of the \ac{LS} correlation estimate and then applies the \ac{GAE} method on the subspace basis vectors, with a complexity that does not increase with the number of observations.   \ac{GEC}  instead performs a separate \ac{GAE} estimation on each new interference observation and then fuses the estimates by clustering the estimated \acp{AoA}. In this case, the complexity increases with the number of observations, since a larger number of points must be clustered; however, the resulting complexity is still less than that of the original \ac{GAE} technique. For comparison purposes, we also consider the \ac{MUSIC} algorithm for the \ac{AoA} estimation.

The rest of the paper is organized as follows. Section~\ref{sysmod} introduces the considered interference scenario, the \ac{mmWave} channel model with few reflections, and the \ac{LS} estimate of the interference correlation matrix. Section~\ref{correst} describes the proposed projection-based correlation estimate and the procedure that first estimates the \acp{AoA} and then provides a more accurate estimate of the interference correlation matrix. The channel interference matrix estimators based on \ac{GAE} techniques are described in Section~\ref{aoaest}. The proposed approaches for the interference correlation estimate are then compared with the \ac{LS} estimate in a typical cellular communication scenario in Section~\ref{numres} before the main conclusions are outlined in Section~\ref{conc}.

\paragraph*{Notation} Matrices and vectors are denoted in boldface, with uppercase and lowercase letters, respectively. $\bm{A}^T$ and $\bm{A}^{\rm H}$ denote the transpose and Hermitian operators on matrix $\bm{A}$, respectively. $\bm{A} \succcurlyeq 0$ indicates that matrix $\bm{A}$ is semidefinite positive. $\bm{I}_L$ is the identity matrix of size $L \times L$. $\bm{B} = {\rm diag}\{\bm{b}\}$ is a diagonal matrix having on the diagonal elements of vector $\bm{b}$. ${\rm atan2}(\bm{p})$ is the 2-argument arc-tangent of point $\bm{p}= [x,y]$. $\bm{A}^{\dag}$ denotes the More-Penrose pseudoinverse of matrix $\bm{A}$.

\section{System Model}\label{sysmod}
 
\revitext{rev21}{In a communication system, we consider one \ac{BS} with $N$ antennas (denoted {\em main} \ac{BS}) receiving signals in uplink from several served \acp{UE}. In the surroundings, a set of users transmit to {\em their own} serving \acp{BS} and interfere at the main \ac{BS}. In this system, we set $L$ as the total number of interferers to the {\em main} \ac{BS}, each interferer is equipped with $N_I$ antennas, and  interferers transmit statistically independent signals.We consider one \ac{BS} (denoted {\em main} \ac{BS}) of  a cellular communication system receiving signals in uplink from several served \acp{UE} through its $N$ antennas.
This scenario fits cellular systems dynamic \ac{TDD} networks (where interferes are in other cells) and cell-free networks.}

Let us indicate with $\bm{J}^{(\ell)}(t) \in \mathbb{C}^{N_I \times 1}$, $\ell=1, \ldots, L$,  the column vector of symbols transmitted by interferer $\ell$ at symbol time $t$.  Assuming time-invariant narrowband channels,  the matrices of the channels from the $L$ interferers to the main \ac{BS} are  $\bm{G}(\ell) \in \mathbb{C}^{N \times N_I}$, $\ell=1, \ldots, L$.    We consider here that only a few interferers are present, i.e., $L$ is a small number. \revitext{rev13-2}{Due to the specific propagation conditions in mmWave, such as severe large-scale fading and blockage, we consider here that only a few interferers are present, i.e., $L$ is a small number.
} \revitext{rev21-2}{The baseband equivalent interference and noise signal received at sample time $t$ by the main \ac{BS} can then be written as the $N$-size column vector}
\begin{equation}\label{eq:1}
 \bm{N}(t)= \bm{Y}(t) + \bm{Z}(t) = \sum_{\ell=1}^{L} \bm{G}(\ell) \bm{J}^{(\ell)}(t)+\bm{Z}(t),
\end{equation}
where $\bm{Z}(t)$  is the vector of \ac{AWGN}, with independent entries having zero mean and variance $\sigma^2$, which is supposed to be known at the main \ac{BS}.  

We assume that  $\bm{J}^{(\ell)}(t)$ and $\bm{Z}(t)$ are independent complex Gaussian vectors with zero mean. The cross-correlation matrices of vectors $\bm{J}^{(\ell)}(t)$ and $\bm{Z}(t)$ are  $\bm{R}_{J}=\mathbb{E}[\bm{J}^{(\ell)}(t)\bm{J}^{(\ell)H}(t)]$, and $\bm{R}_{Z}=\mathbb{E}[\bm{Z}(t)\bm{Z}^{\rm H}(t)] = \sigma^2 \bm{I}_N$, respectively. \revitext{revRJ}{Note that the cross-correlation matrix of $\bm{J}^{(\ell)}(t)$, in general, is not an identity matrix as each interferer may use a transmit beamforming to convey one data stream with its multiple antennas.} The cross-correlation of $\bm{N}(t)$ is therefore
\begin{equation}\label{eq:interference1}
\begin{split}
\bm{R}=&\mathbb{E}\left[\bm{N}(t)\bm{N}^{\rm H}(t)\right]\\  
=& \mathbb{E}\left[\left(\sum_{\ell=1}^{L}\bm{G}(\ell)\bm{J}^{(\ell)}(t) + \bm{Z}(t)\right) \times \right.\\ 
& \left.\left(\sum_{\ell=1}^{L}\bm{G}(\ell)\bm{J}^{(\ell)}(t)+\bm{Z}(t)\right)^\mathrm{H}\right]\\ 
= &\sum_{\ell=1}^{L} \bm{G}(\ell)\bm{R}_{J}\bm{G}^{\rm H}(\ell)+\sigma^2\bm{I}_N.
\end{split}
\end{equation}

In this paper, we aim at estimating matrix $\bm{R}$ from $T$ samples of the interference and noise signal, i.e., from $\bm{N}(t)$,  $t=1,\ldots, T$. \revitext{rev21-3}{The main \ac{BS} can obtain such samples for example in correspondence of pilot signals transmitted by the \ac{UE} canceling the received signals  (and also some other strong interference signals).} The estimated interference correlation matrix $\hat{\bm{R}}$ will then be used to process the forthcoming received signals containing uplink data and mitigate the effects of interference. \revitext{intasnoise2}{As mentioned in the introduction, we consider a scenario wherein the interference is treated as additional noise, as an example in the presence of several low-power interferes whose signals cannot be properly detected and canceled.}

\subsection{Channel Model}

All devices are equipped with linear antenna arrays. Signals transmitted by interferer $\ell$, with $\ell= 1, \ldots, L$, propagate through $N_g$ scatterers and we indicate with $\alpha_{i,\ell}^{\rm (Tx)}$ and $\alpha_{i,\ell}^{\rm (Rx)}$ the \ac{AoD} and \ac{AoA} of the $i$-reflection, for $i=0, \ldots, N_g-1$. Let us also define the transmit and receive phase shifts as
\begin{equation}\label{eq:angles}
\begin{aligned}
\gamma_{i,\ell} =\frac{D_{\min}^{(\mathrm{Tx})}}{\lambda}\sin\alpha_{i,\ell}^{(\mathrm{Tx})},\\
\beta_{i,\ell}=\frac{D_{\min}^{(\mathrm{Rx})}}{\lambda}\sin\alpha_{i,\ell}^{(\mathrm{Rx})},
\end{aligned}
\end{equation}
where $D_{\min}^{(\mathrm{Tx})}$ and $D_{\min}^{(\mathrm{Rx})}$ are the distances between the elements in the transmit and receive linear antenna arrays, respectively, and $\lambda$ is the wavelength of the carrier signal. Defining the steering vector function  
\begin{equation}\label{eq:gridless2}
\bm{a}_{N}(\beta)=[1, e^{-2 \pi {\tt i} \beta }, \ldots, e^{-2 \pi {\tt i} \beta (N-1)}]^{\mathrm{T}},
\end{equation}
with ${\tt i}=\sqrt{-1}$, we can write the interference channel matrices as 
\begin{equation}\label{eq:gridless1}
\bm{G}(\ell)=\sum_{i=0}^{N_g-1} v_{i,\ell} \bm{a}_{N}(\beta_{i,\ell}) \bm{a}_{N_I}^{\mathrm{H}}(\gamma_{i,\ell}),
\end{equation}
where  $v_{i,\ell}$ is the complex gain of signal $i$  of the $\ell$th interferer, with power given by the path-loss ($P_{i, \ell}$). In \ac{mmWave}, the attenuation is high, due to the high operating frequencies and only a few rays participate in the formation of the useful signal (typically $N_g \in \{1, 2, 3\}$). In the following, we assume that the \ac{BS} knows the value of $N_g$, although solutions are available to estimate its value \cite{298274, 1164700}. \revitext{blockage1}{Note that our model may easily include the presence of blockage, a typical phenomenon of \acp{mmWave}, by considering a higher path-loss $P_{i,\ell}$, \cite{6785327, benvenuto2021algorithms}.} \revitext{revposinter}{Moreover, we do not have any specific assumption on the distribution of the interferes that will affect in general the statistics of $v_{i,\ell}$, $\beta_{i,\ell}$, and $\gamma_{i,\ell}$.}

\subsection{\ac{LS} Interference Correlation Matrix Estimate}
\label{LSsubsec}
 
A basic estimate of the interference correlation matrix $\bm{R}$ from the $T$ samples is provided by the \ac{LS} estimate (or sample covariance matrix)
\begin{equation}\label{eq:corrInterf4}
\hat{\bm{R}}_{\rm LS}(T)=\frac{1}{T} \sum_{t=1}^{T}\bm{N}(t)\bm{N}^\mathrm{H}(t).
\end{equation}
Although for $T \to \infty$ we have $\hat{\bm{R}}_{\rm LS} \to \bm{R}$,  the convergence of this estimator alone is slow, thus we consider techniques having a faster convergence. We will also integrate the \ac{LS} estimate into advanced solutions that, starting from a still rough estimate, are able to refine it significantly.

\section{Projection-based Correlation Estimate}\label{correst}

In this section, we propose an estimator of the interference correlation matrix that exploits the structure of the interference channel given by \eqref{eq:1}.  From \eqref{eq:gridless1},  we immediately have
\begin{equation}\label{eq:gridless3}
\bm{Y}^{(\ell)}(t) = \bm{G}(\ell)\bm{J}^{(\ell)}(t)=\sum_{i=0}^{N_g-1}\bm{a}_{N}(\beta_{i,\ell}) \underbrace{[v_{i,\ell} \bm{a}_{N_I}^{\mathrm{H}}(\gamma_{i,\ell})\bm{J}^{(\ell)}(t)]}_{\substack{x_{i,\ell}}(t)},
\end{equation}
where (for fixed $v_{i,\ell}$) $\{x_{i,\ell}(t)\}$, $i=0, \ldots, N_g-1$, are zero-mean complex Gaussian variables (since $\bm{J}^{(\ell)}$ is  assumed to be Gaussian)  with cross-correlations  
\begin{equation}\label{eq:gridless4}
\begin{aligned}
\mathbb{E}[x_{i,\ell}(t) x_{j,\ell}^*(t)]  &=  v_{i,\ell} v_{j,\ell}^* \mathbb{E}\left[\bm{a}_{N_I}^\mathrm{H}(\gamma_{i,\ell})\bm{J}^{(\ell)}(t) \bm{J}^{(\ell)H}(t)\bm{a}_{N_I}(\gamma_{j,\ell})\right] \\ 
&=  v_{i,\ell} v_{j,\ell}^* \bm{a}_{N_I}^\mathrm{T}(\gamma_{i,\ell}) \bm{R}_J \bm{a}_{N_I}\gamma_{j,\ell}).
\end{aligned}
\end{equation} 
Therefore, from \eqref{eq:1} and \eqref{eq:gridless3} we have
\begin{equation}\label{updateY}
\bm{Y}(t) = \sum_{i=0}^{N_g-1} \sum_{\ell=1}^L \bm{a}_N(\beta_{i,\ell}) x_{i,\ell}(t).
\end{equation}
Note that we assume that both $L$ and $N_g$ are small numbers and in particular $N_g L << N$. 

To obtain a more compact model, we collect the interference channels into the matrix  $\bm{G}=[\bm{G}(1), \ldots,\bm{G}(L)]$ and analogously we collect the transmitted interference symbol vectors into the column vector
$\bm{J}(t)=[\bm{J}^{(1) \mathrm{T}}(t), \ldots, \bm{J}^{(L) \mathrm{T}}(t)]^T$. Now, by  defining 
\begin{equation}
\begin{split}
\bm{x}(\ell,t) = [x_{1,\ell}(t), x_{2,\ell}(t), \ldots, x_{N_g-1,\ell}(t)]^T,  \\
\bm{x}(t) = [\bm{x}^T(1,t), \ldots, \bm{x}^T(L,t)]^T,
\end{split}
\end{equation}
we can rewrite the interference vector at sample time $t$ as
\begin{equation}\label{eq:gridless6}
\bm{N}(t)=\bm{G}\bm{J}(t)+\bm{Z}(t)=\bm{A}\bm{x}(t) + \bm{Z}(t),
\end{equation}
with  $\bm{A}= [\bm{A}(1), \ldots, \bm{A}(L)]$ and $\bm{A}(\ell)= [\bm{a}_{N}(\beta_{0,\ell}), \ldots, \bm{a}_{N}(\beta_{N_g-1,\ell})]$. From \eqref{eq:gridless6} we can write the interference  vector as the  sum of $LN_g$ correlated Gaussian sources,  arriving at the BS with receive phase shifts $\beta_{i,\ell}$, for $i=0, \ldots, N_g-1$, and $\ell= 1,\ldots, L$.

Thus, inserting \eqref{eq:gridless6} into \eqref{eq:interference1}, the interference correlation matrix can be written as
\begin{equation} 
\bm{R}=\bm{A}\bm{R}_{\bm{x}}\bm{A}^{\mathrm{H}}+\sigma^2 \bm{I}_N,
\label{decompR}
\end{equation}
where $\bm{R}_{\bm{x}} = \mathbb{E}[\bm{x}(t)\bm{x}^{\mathrm{H}}(t)]$ is block-diagonal with $L$ blocks, as we assumed that the signals transmitted by the $L$ interferers are independent; entries of block $\ell$ are given by \eqref{eq:gridless4}.  From \eqref{decompR}, we observe that the interference correlation matrix (neglecting noise) has a particular structure, as it can be decomposed into the product of three matrices: the two external matrices depend only on the \acp{AoA} of the interference signals, while the inner matrix depends only on the correlation of the interference signals coming from the different angles. 

As we consider a scenario with few paths and few interferers,  the number of rows (columns) $N_gL$ of $\bm{R}_{\bm{x}}$ is much smaller than the number of rows (columns) $N$ of $\bm{R}$. Moreover, although matrix $\bm{A}$ may have a large number of rows ($N$), it actually depends again on a small number of parameters. Therefore, we propose the \ac{PBCE} technique that splits the correlation estimation into two sub-problems: a) the estimation of the receive phase shifts (from which an estimate of matrix $\bm{A}$ is obtained), and b) the estimate of the inner correlation matrix $\bm{R}_{\bm{x}}$.

In particular,  \ac{PBCE}  works as follows. Once $T$ received vectors $\bm{N}(t)$, $t=1,\ldots, T$, have been collected:
\begin{enumerate}
	\item estimate the receive phase shifts as $\hat{\beta}_{i,\ell}$, $i = 0, \ldots, N_g-1$, $\ell=1, \ldots, L$, from  $\bm{N}(t)$, $t=1,\ldots, T$;
	\item obtain the LS estimate of the interference correlation matrix $\hat{\bm{R}}_{\rm LS}(T)$ as in \eqref{eq:corrInterf4};
	\item remove the noise contribution from the LS estimate of the correlation matrix to obtain an estimate of the interference correlation
	\begin{equation}
	\hat{\bm{R}}_{\rm LS}'(T) = \hat{\bm{R}}_{\rm LS}(T) -  \sigma^2 \bm{I}_N; 
	\label{RNL}
	\end{equation}
	\item project $\hat{\bm{R}}_{\rm LS}'(T)$ into the subspace defined by $\hat{\bm{A}}(\ell)= [\bm{a}_{N}(\hat{\beta}_{0,\ell}), \ldots, \bm{a}_{N}(\hat{\beta}_{N_g-1,\ell})]$ and $\hat{\bm{A}}= [\hat{\bm{A}}(1), \ldots, \hat{\bm{A}}(L)]$ to obtain an estimate of $\bm{R}_{\bm{x}}$ as
\begin{equation}\label{eq:gridless9.1}
\hat{\bm{R}}_{\bm{x}}(T) = \hat{\bm{A}}^{\dag}\hat{\bm{R}}'_{\rm LS}(T)(\hat{\bm{A}}^{\dag})^{\mathrm{H}};
\end{equation}
\item  obtain the new estimate of the interference correlation matrix as (see \eqref{decompR})
\begin{equation}
\hat{\bm{R}}_{\rm PBCE}(T) = \hat{\bm{A}}\hat{\bm{R}}_{\bm{x}}(T)\hat{\bm{A}}^{\mathrm{H}}+\sigma^2 \bm{I}_N.
\end{equation}
\end{enumerate}
This is the explanation of the various points:

Point 1) refers to the estimate of the phase shifts and will be detailed in the next Section~\ref{aoaest}, where several techniques will be considered.

Point 2) provides the starting estimate of the interference correlation matrix, from which the inner correlation matrix $\bm{R}_x$ will be estimated, given the estimate of the \ac{AoA} (thus the estimate of $\bm{A}$) obtained in point 1). 

Point 3) elaborates on \eqref{decompR}, first removing the \ac{AWGN} noise contribution from the \ac{LS} estimate of the interference correlation matrix.

Point 4) is obtained by observing that  the \ac{LS} estimate of $\bm{x}(t)$ from $\bm{N}(t)$, given the estimate $\hat{\bm{A}}$, is obtained as (see \eqref{eq:gridless6})
\begin{equation}
\hat{\bm{x}}_{\rm LS}(t) = \hat{\bm{A}}^{\dag} \bm{N}(t).
\end{equation}
Assuming that the angle estimates are correct ($\hat{\bm{A}} = \bm{A}$), the correlation matrix of $\hat{\bm{x}}_{\rm LS}(t)$ is
\begin{equation}\label{Exls}
{\mathbb E}[\hat{\bm{x}}_{\rm LS}(t)\bm{x}_{\rm LS}^{\rm H}(t)] = \bm{A}^{\dag} \bm{R}  \bm{A}^{\dag H}  =  \bm{R}_{\bm{x}} +\sigma^2 \bm{A}^{\dag} \bm{A}^{\dag H},
\end{equation}
and  its \ac{LS} estimate is 
\begin{equation}\label{xLS}
\begin{split}
  \frac{1}{T}\sum_{t=1}^T  &\hat{\bm{x}}_{\rm LS}(t) \hat{\bm{x}}_{\rm LS}^{\rm H}(t)  = \frac{1}{T}\sum_{t=1}^T  \bm{A}^{\dag} \bm{N}(t) \bm{N}^{\rm H}(t) \bm{A}^{\dag H}  \\
  &
  = \bm{A}^{\dag} \hat{\bm{R}}_{\rm LS}(T)  \bm{A}^{\dag H}.
\end{split}
\end{equation}
Then, considering \eqref{xLS} as an estimate of \eqref{Exls} and removing the contribution of noise, we obtain the   estimate  \eqref{eq:gridless9.1} for $\bm{R}_{\bm{x}}$.

Lastly, point 5) follows from \eqref{decompR} and the estimated matrices replace the true matrices. The resulting \ac{PBCE} estimate of $\bm{R}$ can also be written as 
\begin{equation}\label{maxiPBCE}
\hat{\bm{R}}_{\rm PBCE}(T) = \hat{\bm{A}}  \left[\hat{\bm{A}}^{\dag} \hat{\bm{R}}_{\rm LS}(T) \hat{\bm{A}}^{\dag H} - \sigma^2 \hat{\bm{A}}^\dag \hat{\bm{A}}^{\dag H}\right] \hat{\bm{A}}^{\rm H} + \sigma^2 \bm{I}_N.
\end{equation}

\subsection{MSE For Correct Phase Shift Estimates}\label{analysis}

We now compute the \ac{MSE} of the interference correlation matrix estimate under the hypothesis of correct receive phase-shift estimates and compare it with that obtained with the \ac{LS} estimate.  The \ac{MSE} is defined as 
\begin{equation}\label{MSEdef}
\Gamma=\frac{1}{N^2} \sum_{m,n} {\mathbb E}[|\hat{\bm{R}}_{n,m} - \bm{R}_{n,m}|^2].
\end{equation}
First, note that both the \ac{LS} and the \ac{PBCE} estimates (under correct angle estimation) are unbiased, i.e., $\mathbb{E}[\hat{\bm{R}}_{\rm LS}] = \mathbb{E}[\hat{\bm{R}}_{\rm PBCE}] = \bm{R}$.
Then, for the \ac{LS} estimate we have  
\begin{equation}
\Gamma_{\rm LS}(T) =  \frac{1}{N^2} \sum_{m,n} {\mathbb E}\left[\left|\frac{1}{T}\sum_{t=1}^T[\bm{N}(t)  \bm{N}^\mathrm{H}(t)]_{n,m} -  [\bm{R}]_{n,m}\right|^2\right]. \end{equation}
Now, considering that $[\bm{N}(t)  \bm{N}^\mathrm{H}(t)]_{n,m}$ are zero-mean and independent in $t$, we have
\begin{equation}
 \Gamma_{\rm LS}(T) =  \frac{1}{N^2} \sum_{m,n} {\mathbb E}[|[ \bm{N}(t)  \bm{N}^\mathrm{H}(t)]_{n,m} -  [\bm{R}]_{n,m}|^2],
\end{equation} 
and from the results of the Appendix, we have
\begin{equation} 
\begin{split}
 \Gamma_{\rm LS}(T) =&  \frac{1}{TN^2} \sum_{m,n} [\bm{R}]_{n,n}[\bm{R}]_{m,m} \\
= &  \frac{1}{TN^2}{\rm trace}^2(\bm{R}).
\end{split}
\end{equation}

For the \ac{PBCE},  from   \eqref{maxiPBCE}  and assuming $\hat{\bm{A}} = \bm{A}$, we have
\begin{equation}
\begin{split}
\Gamma_{\rm PBCE}(T) = &\frac{1}{N^2} \sum_{m,n}
    {\mathbb E}[|[\bm{R}_{\rm PBCE}]_{n,m} - [\bm{R}]_{n,m}|^2] \\
= &\frac{1}{N^2} \sum_{m,n}
    {\mathbb E}\left[\left|\frac{1}{T}\sum_{t=1}^T \left[\bm{A}  \left[\bm{A}^{\dag} \bm{N}(t)\bm{N}^{\rm H}(t) \bm{A}^{\dag H} \right.\right.\right.\right.\\
    & \left.\left.\left.\left.- \sigma^2 \bm{A}^\dag \bm{A}^{\dag H}\right] \bm{A}^{\rm H} + \sigma^2 \bm{I}_N\right]_{n,m} - [\bm{R}]_{n,m}\right|^2\right]. 
\end{split}
\end{equation}
Since each term of the summation is zero-mean and independent also in this case, we have  
\begin{equation}
\begin{split}
\Gamma_{\rm PBCE}&(T) = \frac{1}{TN^2} \sum_{m,n}
    {\mathbb E}\Big[\Big|\left[\bm{A} (\bm{A}^{\dag} \bm{N}(t)\bm{N}^{\rm H}(t) \bm{A}^{\dag H} -  \right.\\
    &\left.\left.\left.\sigma^2 \bm{A}^\dag \bm{A}^{\dag H}) \bm{A}^{\rm H} + \sigma^2 \bm{I}_N\right]_{n,m} - [\bm{R}]_{n,m}\right|^2\right] \\
    = & \frac{1}{TN^2} \sum_{m,n}
    {\mathbb E}\Big[\left|[\bm{A} \bm{A}^{\dag} \bm{N}(t)\bm{N}^{\rm H}(t) \bm{A}^{\dag H}\bm{A}^{\rm H}]_{n,m}  - \right.\\
    &\left.\left.\left[\sigma^2  \bm{A}\bm{A}^\dag \bm{A}^{\dag H}  \bm{A}^{\rm H} - \sigma^2 \bm{I}_N+ \bm{R}\right]_{n,m}
    \right|^2\right]
\end{split}
\end{equation}
and again from the results of the Appendix, we have
\begin{equation}
\begin{split}
\Gamma_{\rm PBCE}(T) = & \frac{1}{TN^2}\sum_{m,n} [\bm{A} \bm{A}^{\dag} \bm{R} \bm{A}^{\dag H}\bm{A}^{\rm H}]_{n,n} [\bm{A} \bm{A}^{\dag} \bm{R} \bm{A}^{\dag H}\bm{A}^{\rm H}]_{m,m} \\
= & \frac{1}{TN^2} {\rm trace}^2 [\bm{A} \bm{A}^{\dag} \bm{R} \bm{A}^{\dag H}\bm{A}^{\rm H}].
\end{split}
\end{equation} 
Since $\bm{A}^\dag \bm{A} = \bm{I}_{LN_g}$  and using \eqref{decompR}, we have
\begin{equation}
\begin{split}
\Gamma_{\rm PBCE}(T) = & \frac{1}{TN^2}{\rm trace}^2[\bm{R} - \sigma^2 (\bm{I}_N - \bm{A} \bm{A}^{\dag} \bm{R} \bm{A}^{\dag H}\bm{A}^{\rm H})].     
\end{split}
\end{equation}

\section{Receive Phase-Shift Estimation}\label{aoaest}

For the estimation of the receive phase shifts $\hat{\beta}_{i,\ell}$, $i=0,\ldots, N_g-1$, $\ell=1, \ldots, L$, we exploit the assumptions that interferers' channels have only a few paths and there are only a few interferers. This yields a sparse Fourier transform of the correlation matrix, i.e., its sparse representation in the {\em angular} domain. We consider three solutions, based on the gridless approach of \cite{9380368}, which has been proven to provide accurate estimates even from a single or a few samples (small $T$). Still, as $T$ grows, its complexity becomes prohibitive, and we propose also two simplified approaches. In particular, first, we propose the \acf{GAE} based on \cite{9380368,9371398}; in the second approach, denoted as \acf{SGE}, the receive phase shifts are estimated from the \ac{LS}-estimated interference correlation matrix; lastly, in the third approach, denoted \acf{GEC}, \ac{GAE} is performed on windows of samples and the estimated phase shifts are clustered.

\subsection{Gridless Angle Estimation}
\label{GAEsubsec}

First, we consider a solution where the \acp{AoA} are not forced to be on a grid but can take continuous values, in what is called a {\em gridless} estimation. In particular, we resort to the approach of \cite{9380368,9371398}, which leverages the equivalence between the computation of the $\ell_0$ \ac{AN} and a rank minimization problem restricted to the set of positive semidefinite canonical multi-level Toeplitz matrices. 

In the \ac{GAE} algorithm, the receive phase shifts are estimated by imposing a sparsity ($\ell_0$ norm) on the estimated vectors $\hat{\bm{Y}}(t)$, $t=1, \ldots, T$. First, we recall that the $\ell_0$ \ac{AN} of vectors $\{\hat{\bm{Y}}(t)\}$ in the atomic set $\mathcal E = \{\bm{a}_{N}(\beta), \beta \in [0, 2\pi) \}$, is defined as 
\begin{equation}\label{l0def}
||\{\hat{\bm{Y}}(t)\}||_{\mathcal E,0} = \inf_{\{\zeta_p\}, \{q_p\}} \left\{P: \hat{\bm{Y}}(t) = \sum_{p=1}^P q_p(t) \bm{a}_N(\zeta_p), \; \forall t \right\}.
\end{equation}
Note that this definition is consistent with \eqref{updateY} when $\{\zeta_p\}$ are the estimated receive phase shifts. Then, we aim at minimizing the \ac{MSE} between $\bm{N}(t)$ and $\hat{\bm{Y}}(t)$, under the sparsity constraint on $\hat{\bm{Y}}(t)$, with $t=1, \ldots, T$, i.e.,
\begin{equation}
\min_{\{\hat{\bm{Y}}(t)\}} \frac{1-\eta}{T} \sum_{t=1}^T ||\bm{N}(t) - \hat{\bm{Y}}(t)||^2_2 + \eta  ||\{\hat{\bm{Y}}(t)\}||_{\mathcal E, 0},
\label{probl0}
\end{equation}
where $\eta \in (0,1)$ is a parameter ruling the sparsity of the obtained solution. The solution of \eqref{probl0} provides $\hat{\bm{Y}}(t)$, $t=1, \ldots, T$, which in turn provide the estimated receive phase shifts  $\hat{\beta}_{i,\ell}$, $i=0,\ldots, N_g-1$, $\ell=1, \ldots, L$, through \eqref{l0def}.

Now, the function \eqref{probl0} is non-convex, making its minimization hard. Therefore, we consider its $l_1$-AN relaxation to obtain a simpler problem. Let $\mathcal T_{\rm N}$ be the set of positive semidefinite canonical $1$-Level Toeplitz matrices with dimension $N$ (see \cite{9380368} for its definition and properties). The relaxation of \eqref{probl0} leads to the following problem \cite{9380368} 
\begin{subequations}
\begin{equation}
\min_{\tau, \{\bm{q}(t)\}, \bm{Q} \in \mathcal T_{\rm N}} \frac{1-\eta}{T} \sum_{t=1}^T ||\bm{N}(t) - \hat{\bm{Y}}(t)||^2_2 + \frac{\eta}{2}[\tau + {\rm trace}(\bm{Q})],
\end{equation}
s.t.
\begin{equation}
\left[\begin{matrix}
\bm{Q} & \hat{\bm{Y}}(t) \\
\hat{\bm{Y}}^{\rm H}(t) & \tau
\end{matrix}\right] \succcurlyeq 0,\quad t=1, \ldots, T.
\end{equation}
\label{prob1}
\end{subequations}
This is a convex problem that can be solved with standard methods. The new estimates of the receive phase shifts $\hat{\beta}_{i,\ell}$, $i=0,\ldots, N_g-1$, $\ell=1, \ldots, L$,  are then obtained by the Vandermonde decomposition of matrix $\bm{Q}$, as from the \cite[Algorithm 1]{9380368} (see also \cite[Lemma 1]{9371398}).  

\paragraph*{Complexity Analysis} \revitext{comp1}{As shown in \cite{9380368, Krishnan2005}, solving \eqref{prob1} requires} 
\begin{equation}\label{compGAE}
{\mathcal N}(N,T,\epsilon) = {\mathcal O}(T(N^3 +TN^2 + T^2)\sqrt{N}\log(1/\epsilon)),
\end{equation}
\revitext{comp1b}{arithmetic operations, where $\epsilon$ is the accuracy parameter. We note that as $T$ grows, the number of constraints in \eqref{prob1} also grows linearly with $T$, and the complexity of solving the optimization problem grows as ${\mathcal O}(T^3)$, as shown in \cite{9380368}. Hence, we now consider two simplified solutions that are adequate when the number of samples grows large.}

\subsection{Subspace Gridless Estimation}
\label{SGEsubsec}

The second solution is based on the estimation of the receive phase shifts from the \ac{LS} estimate of the interference matrix, rather than on samples $\bm{N}(t)$, $t=1, \ldots, T$. This approach holds  when the number of interfering signals is smaller than the number of \ac{BS} antennas, i.e.,
\begin{equation}
S = LN_g < N.
\label{lesscond}
\end{equation}

First, we observe that we can remove the contribution of the noise by subtracting the noise correlation as in \eqref{RNL}, obtaining $\hat{\bm{R}}_{\rm LS}'(T) = \hat{\bm{R}}_{\rm LS}(T) - \sigma^2\bm{I}_N$. Then, we can improve the LS estimate by imposing the Hermitian symmetry, obtaining
\begin{equation}\label{hermforce}
\hat{\bm{R}}_{\rm LS}'(T) =  \frac{1}{2} \left[ \hat{\bm{R}}_{\rm LS}'(T) + \hat{\bm{R}}_{\rm LS}^{'H}(T)\right].
\end{equation}
Now, from \eqref{decompR}, we observe that the columns of the true interference-only correlation matrix  
\begin{equation}
\bm{R}' =  \bm{R} - \sigma\bm{I}_N
\end{equation}
lay in a sub-space of size $S$. Therefore, we can decompose the LS estimate of the interference-only correlation by \ac{SVD} as
\begin{equation}\label{SVDRLS1}
\hat{\bm{R}}_{\rm LS}'(T) = \bm{V}\bm{\Lambda}\bm{V}^{\rm H},
\end{equation}
where $\bm{V}$ is unitary and $\bm{\Lambda}$ is diagonal with real non-negative entries $\lambda_1 \geq  \ldots \geq \lambda_{N}$, denoted as singular values.

To partially remove the estimation error, we define the tall $N \times S$ sub-matrix $\bm{\Lambda}'$  of $\bm{\Lambda}$ containing the   $S$ largest singular values, i.e.,  $[\bm{\Lambda}']_{s,s} = \lambda_s$, while all other entries of $\bm{\Lambda}'$ are zero, and compute the new correlation estimate 
\begin{equation}\label{RLS22}
\hat{\bm{R}}_{\rm LS}''(T) = \bm{V}\bm{\Lambda}'\bm{V}^{\rm H}.
\end{equation}

Until now we have reduced the estimation error in the \ac{LS} estimate by exploiting the properties of the \ac{SVD} of the interference correlation matrix. Now, by comparing \eqref{RLS22} with \eqref{decompR}, we note that in the columns of $\bm{V}$ we have the steering vectors with the receive phase shifts. 
Thus, the \ac{SGE} algorithm exploits this property. To this end, we first define the {\em square root} matrix of $\hat{\bm{R}}_{\rm LS}''(T)$ as
\begin{equation}
\bm{S} = \bm{V}{\rm diag}\{\sqrt{\lambda_1}, \ldots, \sqrt{\lambda_{S}}\},
\end{equation}
and then we apply the \ac{GAE} method on the columns of $\bm{S}$. In particular, we solve \eqref{prob1} with $T=S$ and $\bm{N}(t)$ replaced by the columns of $\bm{S}$. 

\paragraph*{Complexity Analysis} \revitext{comp2}{The number of operations required by this approach is}  
\begin{equation}\label{compSGE}
{\mathcal N}(N,S,\epsilon) = {\mathcal O}(N^3 + S(N^3 +SN^2 + S^2)\sqrt{N}\log(1/\epsilon)),
\end{equation}
\revitext{comp2b}{where, with respect to \eqref{compGAE} we added also the operations required by the \ac{SVD} ($N^3$). Note that, in this case, regardless of the value of $T$, the number of constraints is always $S$, thus limiting the computational complexity of this solution.}
\subsection{Gridless Estimation and Clustering}
\label{GECsubsec}

We propose also the \ac{GEC} algorithm, wherein \ac{GAE} is performed on windows of $T_0$ samples, and the receive phase shifts are estimated on several windows. They are then suitably clustered and the new receive phase shift estimates are the cluster heads. The obtained solution is denoted as \ac{GEC}.

In detail, let us first define the subset of observations obtained on window $n$ of $T_0$ samples as
\begin{equation}
\mathcal{N}(n) = \{\bm{N}(t), t = nT_0+1, \ldots, nT_0 + T_0\}.
\end{equation}

Then, the \ac{GEC} algorithm works as follows. At time $nT_0$, for $n\geq 1$:
\begin{enumerate}
	\item apply the \ac{GAE} on $\mathcal{N}(n)$ to obtain the estimated  $LN_g$ receive phase shifts $\{\hat{\beta}_{1,1}(n), \ldots, \hat{\beta}_{N_g-1, L}(n)\}$; 
	\item cluster the $n$  previously estimated receive phase shifts $\{\hat{\beta}_{1,1}(1), \ldots, \hat{\beta}_{N_g-1,L}(n)\}$ into $S$ clusters, as detailed in the following;
	\item obtain the estimate of the receive phase shifts as the centroids of the $S$ clusters.
\end{enumerate}

We now detail the clustering procedure and the computation of the centroids. First, note that the k-means algorithm \cite{benvenuto2021algorithms} cannot be immediately applied on the estimated receive phase shifts, since it does not ensure that the centroids are phase shifts in the interval $[0, 2\pi)$. Therefore, we first convert each estimated receive phase shift $\hat{\beta}_{i,\ell}(n)$ into the two-dimensional vector 
\begin{equation}
\hat{\bm{\beta}}_{i,\ell}(n) = [\cos(2\pi\hat{\beta}_{i,\ell}(n)), \sin(2\pi\hat{\beta}_{i,\ell}(n))].
\end{equation}
Hence, the set of all vectors at time $nT_0$ is 
\begin{equation}
\bar{\mathcal B}= \{\hat{\bm{\beta}}_{0,1}(1), \ldots, \hat{\bm{\beta}}_{N_g-1,L}(1), \ldots, \hat{\bm{\beta}}_{0,1}(n), \ldots, \hat{\bm{\beta}}_{N_g-1,L}(n)\}.
\end{equation}
The $S$ clusters are sets $\mathcal{B}_1, \ldots, \mathcal{B}_S$, with $|\mathcal{B}_s| = n/S$: these sets constitute a partition of $\bar{\mathcal B}$, i.e., $\bigcup_s \mathcal B_s = \bar{\mathcal B}$ and ${\mathcal B}_{s_1} \cap {\mathcal B}_{s_2} \mbox{ for $s_1\neq s_2$}$. We use the \ac{MSE} metric for clustering and finding the cluster head
\begin{equation}
\bm{\beta}_s = {\rm argmin}_{\bm{\zeta}} \sum_{\bm{\beta} \in {\mathcal B}_s} ||\bm{\zeta} - \bm{\beta}||^2_2 = \sum_{\bm{\beta} \in {\mathcal B}_s} \bm{\beta}
\end{equation}
Lastly, we convert $\bm{\beta}_s$ into the estimated phase shift as
\begin{equation}
\beta_s = {\rm atan2}( \bm{\beta}_s).
\end{equation}

\paragraph*{Complexity Analysis} \revitext{comp3}{To obtain equal-size clusters of observations, we can resort for example to the algorithm of \cite{sym11030338}, having complexity ${\mathcal O}(T^{1.7})$, thus the number of operations required by this approach is  
\begin{equation}\label{compGEC}
{\mathcal N}(N,T,\epsilon) = {\mathcal O}(T^{1.7}+ T_0(N^3 +T_0N^2 + T_0^2)\sqrt{N}\log(1/\epsilon)).
\end{equation}
}

\subsection{MUSIC Angle Estimation}\label{secMUSIC}

For comparison purposes, we consider the \ac{MUSIC} algorithm \cite{1143830}    to estimate the receive phase shift. This algorithm is based on the \ac{SVD} of the LS estimate; as for \ac{SGE}, we consider the LS-estimated interference-only correlation matrix $\hat{\bm{R}}_{\rm LS}'(T)$ and its \ac{SVD} \eqref{SVDRLS1}.  Then, the receive phase shifts $\{\beta_{i,\ell}\}$ are estimated by finding  the $LN_g$ values of $\theta$ corresponding to the largest peaks of the frequency estimation function
\begin{equation}
P_{\rm MU}(\theta) =  \frac{1}{\bm{a}_N^{\rm H}(\theta)\bm{V}\bm{V}^{\rm H}\bm{a}_N(\theta)}.
\end{equation}
Note that the peak is searched in a continuous space; however, in practice, $\theta$ is sampled in a discrete space over which the maxima are searched, thus a grid approach is obtained. 

\paragraph*{Complexity Analysis} \revitext{comp4}{The complexity of \ac{MUSIC} then depends on the number of points of the grid $N_{\rm G}$, and can be written as \cite{6422415}
\begin{equation}
    {\mathcal O}(N^2(LN_g+T+N_{\rm G})).
\end{equation}
}

\section{Numerical Results}\label{numres}

\begin{figure} 
	\centering
	\includegraphics[width =0.6\hsize]{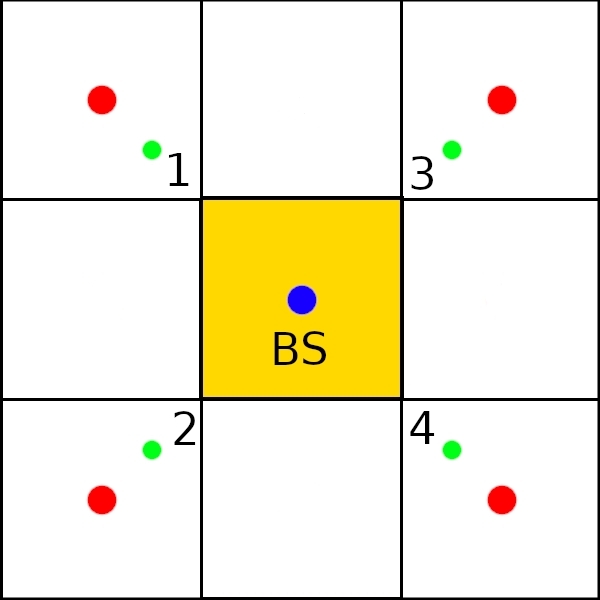} 
	\caption{Interference environment with the main \ac{BS} (blue dot), neighborhood cells (white squares) with their \acp{BS} (red dots), and interferers (green dots).}
	\label{fig:int_env}
\end{figure} 
 
\begin{figure*} 
	\centering
	\begin{tabular}{cc}
	\includegraphics[width =0.5\hsize]{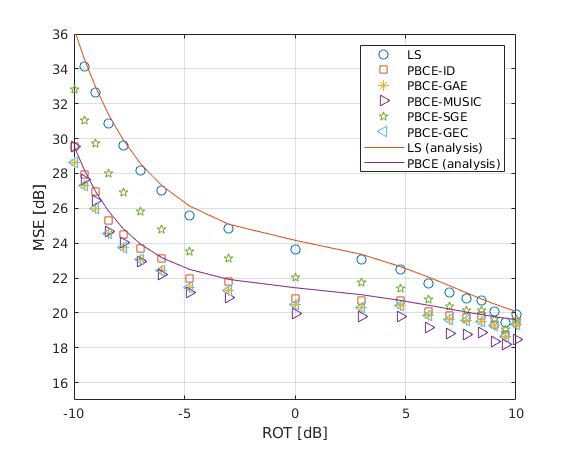} &
	\includegraphics[width =0.5\hsize]{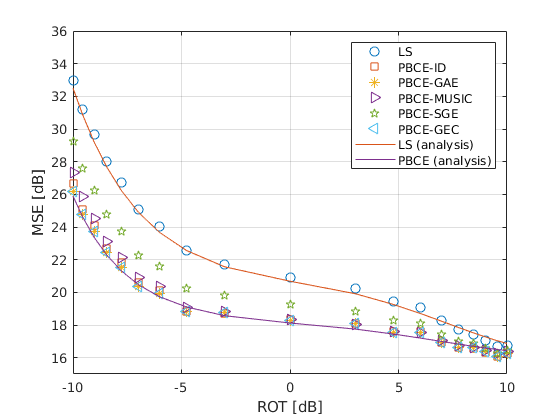} \\ \small
	a) Case $T=2$ samples. & \small b) Case $T=4$ samples.  
	\end{tabular}
	\caption{MSE vs \ac{ROT} for $T=2$ and $T=4$ samples and various interference correlation matrix estimators.}
	\label{fig:MSE}
\end{figure*}

We assess now the performance of the proposed interference correlation matrix in the scenario of  Fig.~\ref{fig:int_env}, where the {\em main} \ac{BS} (blue point) is at the center of its square cell (in yellow), and the cell is surrounded by 4 neighboring cells (in white), in which $L = 4$ interferers (green points)  communicate with their own  secondary \acp{BS} (red points). The interferers are placed along the segments between the main and the secondary \acp{BS}, which define the average \ac{AoA} for channels $\bm{G}(\ell)$. \revitext{blockage2}{For simulation purposes, we draw ray amplitudes $v_{i,\ell}$ from a complex Gaussian distribution with zero mean and unitary variance, i.e., $P_{i,\ell}=1$, modeling interferers in \ac{nLOS} condition with respect to the main \ac{BS}. Note that such a scenario is the worst-case for interference estimation since all channel paths are weak and thus particularly challenging to be estimated.}

\revitext{pi6}{The main \ac{BS} is equipped with $N=32$ antennas, and the channel is modeled as in Section~II.A, with $N_g = 3$ rays.  \acp{AoA} $\{\alpha^{(\mathrm{Rx})}_{i,\ell}\}$ are uniformly distributed with mean given by the angle between the transmitting interferer and the main BS and support size  $\frac{\pi}{6}$, which corresponds to a cell with 12 sectors (a higher-order sectorization, see \cite{7128401}),  while  \acp{AoD} $\{\alpha^{({\mathrm{Tx}})}_{i,\ell}\}$ are uniformly distributed in the interval $[0,\pi]$.}  The antenna spacings are  $D_{\mathrm{min}}^{(\mathrm{Tx})} = D_{\mathrm{min}}^{(\mathrm{Rx})}=\frac{\lambda}{2}$, for a carrier frequency  $f_c=28$~GHz. 

The performance will be shown as a function of the \ac{ROT}, defined as the ratio between the interference power and the noise power, i.e., 
\begin{equation}\label{ROTeq}
    {\rm ROT} = \frac{\frac{1}{N} {\rm trace} \bm{R}  - \sigma^2}{\sigma^2} = \frac{\frac{1}{N} {\rm trace} \bm{R}}{\sigma^2} -1. 
\end{equation} 
\revitext{revROT}{In particular, in the following we vary the noise power $\sigma^2$ while leaving unchanged the interference power. Therefore, from \eqref{ROTeq}, we note by increasing the noise power we decrease the ROT.}

We focus on the case of single-antenna users and interferes ($N_I=1$) and the  received vector signal at sample $t$ (including also the data signal, not only noise and interference) is
\begin{equation}
\bm{r}(t) = \bm{h}s(t) + \bm{N}(t),
\end{equation}
with $\bm{h}$ the column channel vector between the transmitter in the main cell and the \ac{BS} and  $s(t)$ the transmitted useful data symbol at time $t$, assumed to be Gaussian zero-mean with variance $\sigma_x^2$. We assume here that the \ac{BS} perfectly knows $\bm{h}$. 
 
We assume that the \ac{BS} requests from the transmitter a data rate $R$ to be used for uplink transmission. Since the \ac{BS} does not have a perfect estimate of the interference correlation matrix, its estimate $\hat{C}$ of the achievable rate in the uplink channel is affected by errors. Thus, the \ac{BS} will transmit at rate $R = \delta \hat{C}$, with $\delta \in [0,1]$, suitably chosen, as detailed in the following. Note that even with this choice, {\em outages} may occur, when $R > C$, and $C$ is the true achievable rate of the uplink channel.

\subsection{Performance Metrics}

Performance has been compared in terms of a) \ac{MSE} of the estimate of the interference correlation matrix, b) achievable data rate, and c) effective throughput (taking into account outages). 

\paragraph*{\bf MSE} The \ac{MSE}  of the estimate of the interference correlation matrix is defined in \eqref{MSEdef}.

\paragraph*{\bf Achievable Rate} To define the achievable rate and the throughput, we first revise the operations performed at the \ac{BS} upon reception of a message. \revitext{intasnoise3}{We recall that we treat the interference as additional noise, but we explicitly exploit the knowledge of its correlation matrix, estimated with the techniques presented in this paper. Thus, the \ac{BS} first whitens the interference on $\bm{r}(t)$.} This is achieved by decomposing the estimated interference correlation matrix $\hat{\bm{R}}$ by computing its \ac{SVD}  $\hat{\bm{R}}=\bm{U}\bm{\Lambda}\bm{U}^{\mathrm{H}}$ and then whitening the signal to obtain  
\begin{equation}\label{eq:cd3}
\bm{r}'(t)=\bm{\Lambda}^{-1/2}\bm{U}^{\mathrm{H}}\bm{r}(t)=\bm{g}s(t)+\bm{N}'(t),
\end{equation}
where 
\begin{equation}\label{eqchannel}
\bm{g}=\bm{\Lambda}^{-1/2}\bm{U}^{\mathrm{H}}\bm{h}
\end{equation}
is the equivalent channel and $\bm{N}'(t)$ is the noise term (that includes the noise-like interference). Note that if the estimation of the interference correlation matrix is perfect, i.e., $\hat{\bm{R}} = \bm{R}$, then $\bm{N}'(t)$ has an identity correlation matrix. Otherwise, it is still colored with correlation matrix
\begin{equation}
{\bm{R}}_{\bm{N}'} =  \bm{\Lambda}^{-1/2}\bm{U}^{\mathrm{H}} \bm{R} \bm{U} \bm{\Lambda}^{-1/2}.
\end{equation}
However, the \ac{BS}  is not aware of the residual correlation and cannot exploit it to compute the achievable data rate. 
\revitext{singleUE}{Assuming for the sake of simplicity a single user transmitting in uplink to the main \ac{BS} (i.e., a resource block is assigned to a single user), the \ac{BS} then applies a maximal ratio combiner (MRC)} \revitext{explainC}{to obtain 
\begin{equation}
\bm{r}''(t)= \bm{g}^\mathrm{H}\bm{r}'(t) = \bm{g}^\mathrm{H} \bm{g}s(t)+\bm{g}^\mathrm{H}\bm{N}'(t),
\end{equation}
where now the correlation of the noise becomes $\bm{g}^\mathrm{H}{\bm{R}}_{\bm{N}'}\bm{g}$. Hence,
and the true achievable data rate at the \ac{BS} is
\begin{equation}\label{eq:cd9}
C=\log_2\left(1+\frac{(\bm{g}^\mathrm{H} \bm{g})^2 \sigma_{x}^2}{\bm{g}^\mathrm{H}{\bm{R}}_{\bm{N}'}\bm{g}}\right).
\end{equation}
}

\paragraph*{\bf Throughput} The throughput takes into account outages in the uplink transmission. The  estimated  achievable rate at the \ac{BS} is 
\begin{equation}\label{eq:cd10}
\hat{C}=\log_2\left(1+ \bm{g}^\mathrm{H} \bm{g}   \sigma_{x}^2 \right).
\end{equation}
When $\hat{\bm{R}} = \bm{R}$,  $\hat{C} = C$ and this is also the spectral efficiency of the system, denoted as $C_{\rm opt}$:  it  provides an upper bound to the achievable rate $\hat{C}$. The  throughput  is then defined as
\begin{equation}
\rho = \begin{cases}
0, & \delta\hat{C} > C, \\
\delta\hat{C}, & {\rm otherwise}.
\end{cases}
\end{equation}
We choose $\delta$ to maximize ${\mathbb E}[\rho]$: this is obtained by numerical methods.  

\subsection{Compared Solutions}

We compare the following techniques for the interference correlation matrix:  
\begin{itemize}
    \item \ac{LS}: the \ac{LS} estimate of Section~\ref{LSsubsec};
    \item \ac{PBCE}-\ac{GAE}:  \ac{PBCE}  using the \ac{GAE} algorithm of Section~\ref{GAEsubsec} for the estimate of the receive phase shifts;
    \item \ac{PBCE}-\ac{SGE}:   \ac{PBCE}  using the \ac{SGE} algorithm of Section~\ref{SGEsubsec} for the estimate of the receive phase shifts;
    \item \ac{PBCE}-\ac{GEC}:   \ac{PBCE}   using the \ac{GEC} algorithm of Section~\ref{GECsubsec} for the estimate of the receive phase shifts;
    \item \ac{PBCE}-\ac{MUSIC}:   \ac{PBCE}  using the \ac{MUSIC} algorithm of Section~\ref{secMUSIC} for the estimate of the receive phase shifts;
    \item \ac{PBCE}-\ac{ID}:   \ac{PBCE}   with the perfect  estimate of the receive phase shifts.
\end{itemize}

We have considered $T_0=1$ and $N_{\rm G} = 10^3$.

\subsection{Performance Results} 

First, we compare the performance in terms of \ac{MSE} of the interference correlation  matrix estimate. 

Fig.~\ref{fig:MSE} shows the interference correlation matrix \ac{MSE} \eqref{MSEdef} as a function of the \ac{ROT} for the considered estimators and  two values of the number of observed samples, namely $T=2$ and 4. Note that we consider a very small number of samples to better assess the performance of the estimators based on the \acp{AoA}. In the following, we will also consider larger values of $T$.  We first observe that as the noise increases, the ROT decreases, and the advantage of the proposed techniques becomes more evident since they   distinguish noise from interference better than the \ac{LS} technique. When comparing the various \ac{PBCE} algorithms, we note that the \ac{PBCE}-\ac{SGE} provides only a partial improvement with respect to the \ac{LS} estimator, while other approaches further reduce the \ac{MSE} and achieve similar performance.  Fig.~\ref{fig:MSE} shows also the \ac{MSE} obtained with the analysis of Section~\ref{analysis}, which well matches the simulation results.

\begin{figure*}
	\centering
	\begin{tabular}{cc}
	\adjincludegraphics[width=.5\hsize, ]{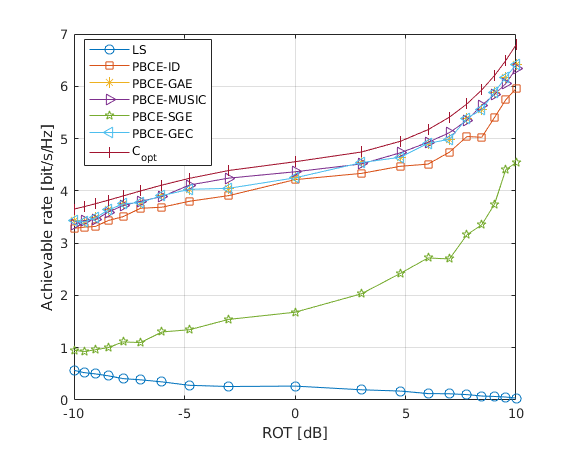}&
	\adjincludegraphics[width=.5\hsize]{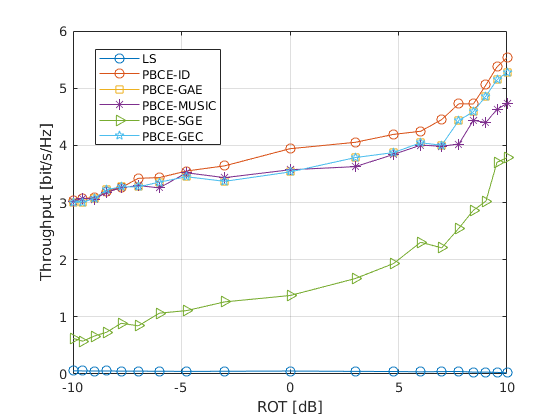}
\\ \small
a) Achievable rate & \small b) Throughput
		\end{tabular}
	\caption{Achievable rate (a) and throughput (b) vs \ac{ROT} using $T=2$ samples  and various interference correlation matrix estimators.}
	\label{fig:capa1}
\end{figure*}

We now consider both the achievable rate and the throughput as performance metrics.

Fig.~\ref{fig:capa1} shows both the achievable rate and the throughput as a function of the \ac{ROT} for $T=2$ samples and various correlation estimation methods. We observe that the \ac{LS} estimate for such a small value of $T$ provides a very low achievable rate, much lower than that achieved when using \ac{PBCE} techniques for the interference correlation matrix estimation. Indeed, all \ac{PBCE} methods achieve a rate close to the capacity $C_{\rm opt}$, all with similar performance, except for the \ac{PBCE}-\ac{SGE} method, showing a higher sensitivity to the interference and yielding an achievable rate close to  \ac{LS}  for high \ac{ROT} values. Lastly, note that all \ac{PBCE} techniques exhibit an achievable rate reduction as the \ac{ROT} increases, since the overall interference increases. For the \ac{LS} technique, the estimation is very bad and the slight increase of the rate with the \ac{ROT} is due to the fact that interference is structured and the whitening becomes a bit more effective. When comparing Fig.~\ref{fig:MSE} and Fig.~\ref{fig:capa1}, we conclude that the \ac{MSE} of the correlation matrix estimate is not indicative of the data rate performance, as the correlation matrix is used for whitening (see \eqref{eqchannel} and \eqref{eq:cd9}) and the impact of correlation matrix estimation errors is not linear on the data rates.


\begin{figure}
	\centering
	\includegraphics[width =0.6\hsize,min width =8.5cm]{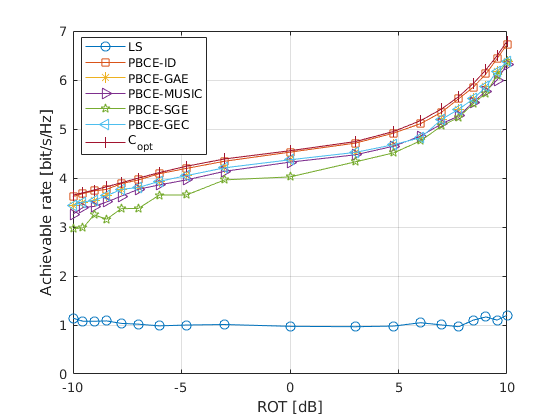}
	\caption{Achievable rate vs \ac{ROT} with $T=5$ samples  and various interference correlation matrix estimators.}
	\label{fig:capa5}
\end{figure}

Fig.~\ref{fig:capa5} shows the achievable rate as a function of the \ac{ROT}, when $T=5$~samples, for various correlation estimation methods. The trend is similar to that of Fig.~\ref{fig:capa1}, with the \ac{PBCE} solutions getting closer to the upper bound, and the \ac{PBCE}-\ac{SGE} approach significantly increasing the achievable rate and aligning with the other techniques. Even the \ac{LS} approach yields a higher achievable rate but is still considerably reduced with respect to that of the \ac{PBCE} techniques. The \ac{LS} method is still significantly affected by the estimation error, providing a very low achievable rate: indeed, the estimation error is so dominant that it turns out to be almost insensitive to the \ac{ROT}.

\begin{figure} 
	\centering
	\includegraphics[width =0.6\hsize,min width =8.5cm]{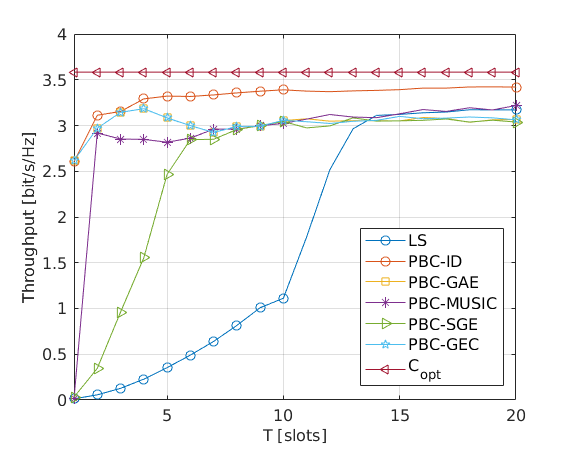}
	\caption{Throughput vs $T$ with $ROT=-10$~dB  and various interference correlation matrix estimators.}
	\label{fig:thr10}
\end{figure}

\begin{figure} 
	\centering
	\includegraphics[width =0.6\hsize,min width =8.5cm]{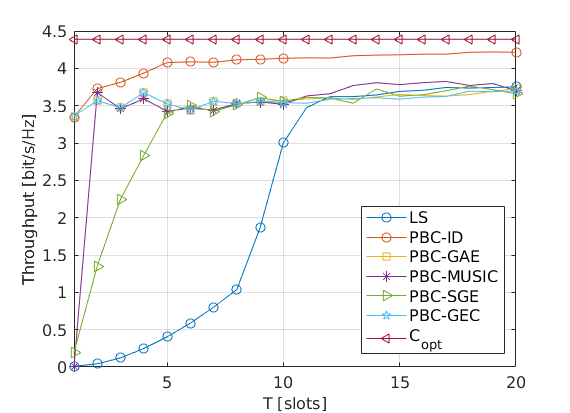}
	\caption{Throughput vs $T$ with $ROT=0$~dB  and various interference correlation matrix estimators.}
	\label{fig:thr0}
\end{figure}

Lastly, Figs.~\ref{fig:thr10} and~\ref{fig:thr0} show the throughput as a function of the number of samples $T$, for $ROT=-10$ and 0~dB, respectively. As expected, a larger number of samples used for the estimation provides higher throughput. However, the \ac{PBCE} approaches show a much faster increase of the throughput with the number of samples than the \ac{LS} approach, which is a significant advantage of the proposed solutions.  Indeed, the \ac{LS} approach suffers from the increased interference and requires more samples to achieve values close to the optimal when the ROT increases, and even when using more than 10 samples, the throughput improvement is very slow. Note also that for high values of ROT, the \ac{PBCE} method based on \ac{GEC} outperforms \ac{PBCE}-\ac{MUSIC} providing a 10\% increase in the throughput.

\revitext{rev3CS}{In summary, we note that the proposed \ac{GEC} methods outperform both the \ac{LS} and the \ac{MUSIC} techniques, at least at certain values of \ac{ROT}. Moreover, \ac{GEC} offer a faster convergence than \ac{MUSIC}.}

\subsection{Complexity Comparison}

\begin{figure} 
	\centering
	\includegraphics[width =0.6\hsize,min width =8.5cm]{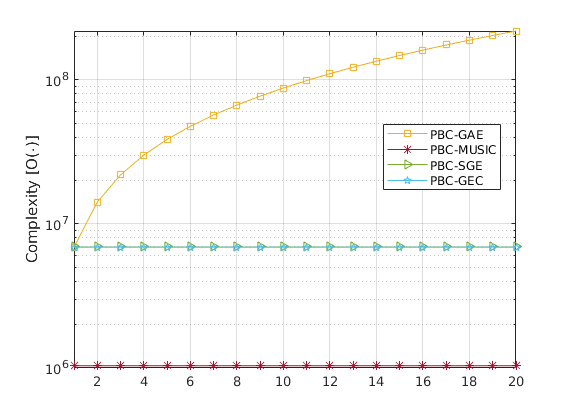}
	\caption{Complexity comparison among the techniques.}
	\label{fig:complex}
\end{figure}

\revitext{compfig}{We have also compared the complexity of the various receive phase-shift estimation algorithms, as shown in Fig.~\ref{fig:complex} that reports the asymptotic ($\mathcal O(\cdot)$) number of operations required as a function of $T$, with all other parameters set as in the other performance figures. We note that \ac{PBCE}-\ac{GAE} has the highest complexity, which also grows fast with $T$, while both \ac{PBCE}-\ac{SGE}  and \ac{PBCE}-\ac{GEC} have a complexity independent of $T$, as indeed it is related to $T_0$ and $S$, which are here kept constant. The \ac{MUSIC} algorithm exhibits the lowest complexity, as expected. From the figure we first appreciate the reduction of complexity achieved by \ac{PBCE}-\ac{SGE}  and \ac{PBCE}-\ac{GEC} with respect to \ac{PBCE}-\ac{GAE}. Then, taking into account the performance (in terms of throughput) achieved by the proposed \ac{PBCE} techniques, we note that this advantage over \ac{MUSIC} comes at the cost of higher complexity.}

\section{Conclusions}~\label{conc}

In this paper, we have proposed three techniques for the estimation of the interference correlation matrix in a  cellular system. The approaches are based on the estimation of the \acp{AoA} of interference signals for channels with few rays. The investigation of several techniques for the estimation of the receive phase shift and their performance comparison in a simulated environment has shown that the \ac{GAE} algorithm is very effective  and that the sub-optimal (and less complex) solution \ac{GEC} is also well-performing, especially when more symbols are available for the interference correlation estimate. \revitext{rev3paper}{A future study related to our proposed estimators may be devoted to the design of interference correlation matrix estimators when hybrid analog-digital receive structures are employed: in this case, the constraints on the number of radio-frequency chains and thus the accessible signals call for specific solutions that exploit both the geometry of the receive antennas and their grouping.}

\bibliographystyle{IEEEtran}
\bibliography{biblio.bib}

\appendix

Let $x$ and $y$ be zero-mean jointly circularly symmetric complex Gaussian variables, with powers $\sigma_x^2$ and $\sigma_y^2$,  cross-correlation ${\mathbb E}[xy^*] = \xi\sigma_x\sigma_y$, and $\xi = \xi_R + j \xi_I$. We now aim at computing the variance of the random variable $xy^*$. 

Variable $y$ can be written as a function of $x$ as 
\begin{equation}
    y = \xi^*\frac{\sigma_y}{\sigma_x} x +   \sqrt{\sigma_y^2 - |\xi\sigma_y|^2} w,
\end{equation}
with $w$ zero-mean unitary-variance circularly symmetric complex Gaussian variable, independent of $x$.

Then the variance of the random variable $xy^*$ is
\begin{equation}
\begin{split}
{\mathbb E}[|xy^*  & - \xi\sigma_x\sigma_y|^2] = {\mathbb E}[|xy^*|^2) + |\xi\sigma_x\sigma_y|^2 +\\
& - 2{\mathbb E}(\Re\{xy^*\xi^*\sigma_x\sigma_y\}].
\end{split}
\end{equation}
Since we have 
\begin{equation}
\begin{split}{\mathbb E}[|xy^*|^2] & = {\mathbb E}[|x|^4] |\xi|^2\frac{\sigma_y^2}{\sigma_x^2} + (\sigma_y^2 - |\xi\sigma_y|^2) \sigma_x^2 \\
& = 2\sigma_x^4 |\xi|^2\frac{\sigma_y^2}{\sigma_x^2} + (\sigma_y^2 - |\xi\sigma_y|^2) \sigma_x^2 ,
\end{split}
\end{equation}
\begin{equation}
\begin{split}
  {\mathbb E}& \left[\Re\{xy^*\xi^*\sigma_x\sigma_y\}\right] =  \sigma_x\sigma_y [\xi_R \Re\{xy^*\} - \xi_I \Im\{xy^*\}] = \\
&  \sigma_x^2\sigma_y^2 |\xi|^2,
\end{split}
\end{equation} 
we obtain
\begin{equation}
\begin{split}
    {\mathbb E} & [|xy^*  - \xi\sigma_x\sigma_y|^2] = 2|\xi|^2\sigma_x^2\sigma_y^2   \\
    & +\sigma_x^2\sigma_y^2 (1-|\xi|^2)-2\sigma_x^2\sigma_y^2|\xi|^2+|\xi|^2\sigma_x^2\sigma_y^2 = \sigma_x^2\sigma_y^2.  
    \end{split}
\end{equation}

\end{document}